\documentclass[manuscript]{acmart}

\AtBeginDocument{%
  \providecommand\BibTeX{{%
    \normalfont B\kern-0.5em{\scshape i\kern-0.25em b}\kern-0.8em\TeX}}}

\usepackage{graphicx}
\usepackage{caption}
\usepackage{subcaption}

\begin{document}

\title[Designing AI for Context Analysis in Humanitarian Frontline Negotiations]{``ChatGPT, Don't Tell Me What to Do'': Designing AI for Context Analysis in Humanitarian Frontline Negotiations}

\author{Zilin Ma}
\email{zilinma@g.harvard.edu}
\orcid{0000-0002-7259-9353}
\affiliation{%
  \institution{Intelligent Interactive Systems Group, Harvard School of Engineering and Applied Sciences}
    \streetaddress{150 Western Ave.}
  \city{Allston}
  \state{MA}
  \country{USA}
  \postcode{02134}
}

\author{Yiyang Mei}
\email{yiyang.mei@emory.edu}
\orcid{0009-0001-2923-0729}
\affiliation{%
  \institution{Law School, Emory University}
    \streetaddress{1301 Clifton Road}
  \city{Atlanta}
  \state{GA}
  \country{USA}
  \postcode{30322}
}

\author{Claude Bruderlein}
\email{cbruderl@hsph.harvard.edu}
\orcid{}
\affiliation{%
  \institution{Harvard Kennedy School of Government, Harvard University}
    \streetaddress{79 John F. Kennedy St}
  \city{Cambridge}
  \state{MA}
  \country{USA}
  \postcode{02138}
}

\author{Krzysztof Z. Gajos}
\email{kgajos@eecs.harvard.edu}
\orcid{0000-0002-1897-9048}
\affiliation{%
  \institution{Intelligent Interactive Systems Group, Harvard School of Engineering and Applied Sciences}
  \streetaddress{150 Western Ave.}
  \city{Allston}
  \state{MA}
  \country{USA}
  \postcode{02134}
}

\author{Weiwei Pan}
\email{weiweipan@g.harvard.edu}
\orcid{}
\affiliation{%
  \institution{Harvard School of Engineering and Applied Sciences}
    \streetaddress{150 Western Ave.}
  \city{Allston}
  \state{MA}
  \country{USA}
  \postcode{02134}
}


\begin{abstract}
Frontline humanitarian negotiators are increasingly exploring ways to use AI tools in their workflows. However, current AI-tools in negotiation primarily focus on outcomes, neglecting crucial aspects of the negotiation process. Through iterative co-design with experienced frontline negotiators (n=32), we found that flexible tools that enable contextualizing cases and exploring options (with associated risks) are more effective than those providing direct recommendations of negotiation strategies. Surprisingly, negotiators demonstrated tolerance for occasional hallucinations and biases of AI. Our findings suggest that the design of AI-assisted negotiation tools should build on practitioners' existing practices, such as weighing different compromises and validating information with peers. This approach leverages negotiators' expertise while enhancing their decision-making capabilities. We call for technologists to learn from and collaborate closely with frontline negotiators, applying these insights to future AI designs and jointly developing professional guidelines for AI use in humanitarian negotiations.


\end{abstract}





\maketitle
\section{Introduction}

Humanitarian negotiations in conflict zones, known as frontline negotiation, are crucial for securing access to crisis-affected populations for aid delivery~\cite{GISF2020, CCHN2019}. These negotiations are often highly adversarial, complex, and high-risk; and the stakeholders involved are diversely distributed in terms of geography, politics, and culture. A key challenge for negotiators is to navigate conflicting perspectives by accurately and quickly synthesizing unstructured information from various sources, including interviews, stakeholder meetings, and historical documents~\citep[e.g.][]{pon2023, bruderlein2023}. As the number of violent conflicts and the need for frontline negotiations continue to grow globally~\cite{csis2024, crisisgroupwatchlist2024, cfr2024, courthousenews2024}, professional organizations and individual practitioners have begun exploring the use of AI to support negotiation in order to alleviate increasing pressure on human experts. So far, these explorations are happening organically within communities of negotiation practitioners, without specialized tools or formal collaboration with machine learning or HCI experts. For example, in 2024, Frontline Associates -- a global network of negotiation practitioners -- organized an educational summer program focused on the impact of AI in negotiation~\cite{frontline_summer_program_2024}. Over a hundred humanitarian frontline negotiators from a wide range of organizations (e.g. Médecins Sans Frontières, International Committee of the Red Cross and United Nations) participated in the program and experimented with using ChatGPT in context analysis and information synthesis.

However, given the high-stakes and sensitive nature of frontline negotiation, as well as the well-documented limitations of LLMs, it is unclear how these models can or should be responsibly and ethically used in these contexts. In particular, it is unlikely that negotiators have the technical training to independently assess the risks of using LLMs in their work. For example, as current literature documented, LLMs summaries may fail to capture important subtext, as LLMs lack contextual language understanding and is poor at modeling conversational empathy~\cite{ma_understanding_2023, ma2024evaluating, shen2024understandingcapabilitieslimitationslarge}; these models often exhibit western bias~\cite{stochastic-parrot, jain2023generativeaiwritingresearch, wang2023humancenteredsocialcenteredartificialintelligence, Ryan2024UnintendedIO}, making the deployment of such models potentially ineffective in non-western contexts~\cite{sutton_empathy_2022}. Confidentiality concerns naturally arise when using proprietary LLM models~\cite{ray2023samsung}, and LLMs' tendency to generate inaccurate information (hallucination) presents challenges in negotiation contexts where precision is crucial~\cite{kuhn2023semantic, farquhar2024detecting}. Finally, chat interfaces are flexible but can be difficult for new users who are not familiar with writing effective prompts~\cite{Zamfirescu2023Johnny}.

Beyond quantifying risks posed by the inherent limitations of LLMs in negotiation, there is also a need to anticipate potential problematic interactions between human practitioners and AI models. 
In AI-assisted decision making, achieving human-AI complementarity is challenging, with combinations often performing worse than either alone~\cite{kamar2016directions, inkpen2023complementarity, kamar2012combining, bansal2021does}. This typically results from inappropriate reliance on AI systems (overreliance and underreliance), leading to reduced overall performance~\cite{bucinca2020, bucinca2021trust, Vasconcelos2023Explanations}. 

Given the above concerns about the potential harms that can result from ad-hoc user-driven adoption of LLMs in frontline negotiation, there is a clear need for careful and systematic need-finding studies for this community, as well as a need for works that thoughtfully map user needs to design decisions. While there is existing research on the application of AI in negotiation, they target negotiation contexts that are not aligned with the constraints and goals of humanitarian negotiation, such as zero-sum settings in contracts negotiation. These current approaches are often goal-centric, focusing on using statistical benchmarks to achieve high accuracy. The emphasis is on outcome rather than the process; and these works largely seek to ``surpass human performance'', potentially replacing human negotiators~\cite{kramar_negotiation_2022, Abdelnabi2024Cooperation, meta_fundamental_ai_research_diplomacy_team_fair_human-level_2022}. 

Frontline negotiation, however, needs to achieve human rapport and requires strategy contextualization. Frontline negotiators need to keep abreast of the emotions, cultural backgrounds, and human irrationality involved in the negotiation process~\cite{CCHN2019, clements_humanitarian_2021, elfenbein2017makes,curhan2022silence}. This focus on the human element aligns with the approaches in human-AI collaboration -- particularly the``Process-Oriented Support''~\cite{zhang2024recommendationsbackwardforwardai}. Process-oriented support is an approach to AI decision support systems that focuses on assisting users throughout the entire decision-making process rather than just providing end-to-end recommendations. This concept is particularly relevant in complex, high-stakes environments, where decision-making involves multiple steps and requires integrating diverse pieces of information~\cite{zhang2024recommendationsbackwardforwardai}. Given the nuance and high stakes involved in frontline negotiation, process-oriented support may be more appropriate than the current goal-centric approach.  

In this work, we systematically studied the decision-support needs that are specific to frontline humanitarian negotiators, and we connected these needs to concrete design principles for AI decision-support in humanitarian negotiation. We conducted two studies: (1) a study where we characterized the current workflows of practitioners in frontline humanitarian negotiations, as well as identified their attitudes towards and usage patterns for LLMs; (2) a study where we compared the impacts of process-oriented decision support and goal-centric decision support on practitioner workflows.

In our first study, we ask:
\begin{itemize}
    \item For what purposes do negotiators use LLMs, and what do they hope to achieve with them?
    \item What concerns do negotiators have when applying LLMs in negotiations?
    \item What challenges do negotiators face when using LLMs in frontline negotiations?
\end{itemize}
To address the questions above, we conducted a formative study with 14 frontline negotiators. We identified the following use of LLMs in frontline negotiation: context analysis, ideation support through alternative option generation, knowledge sharing among the negotiators and organizations, and training new negotiators. 

Based on these insights, we developed a probe interface using Process-Oriented Design principles, focusing on context analysis and ideation support. To further compare and evaluate the difference between ChatGPT and our custom interface, and to ascertain the benefits and limitations of LLMs in negotiation, we conducted a comparative study with 15 experienced negotiators. 
This study not only reaffirmed concerns previously identified in LLM literature (such as confidentiality, hallucination, and bias) but also uncovered new issues specific to frontline negotiation contexts, including mandators' influence, resource constraints of LLMs, and the problem of LLMs' overconfidence. We also found that: 
\begin{itemize}
\item Negotiators question if the added benefit of using ChatGPT outweigh the added human labor and uncertanties caused by LLM's inherent challenges. Currently, negotiators do not see LLM tools becoming essential part of frontline negotiation. 
\item Negotiators found it hard to use ChatGPT in real negotiations because they could not easily break down their process into prompts. Our probe interface helped solve this problem by using chain of thoughts prompting, making it simpler to use LLMs in negotiations. 
\item Negotiators' existing validation practices among negotiators help mitigate the effects of AI biases and hallucinations, making them tolerable. The extent to which these practices mitigate the negative effects is dependent on the negotiators' negotiation expertise. Our probe interface supported these practices, potentially reducing overreliance on LLMs.  
\item Current outcome-focused AI approaches in negotiation mis-align with negotiators' process-oriented needs, which emphasize context understanding and rapport-building.
\end{itemize}

Our findings point to clear design implications for decision-support tools in frontline humanitarian negotiation. Future AI decision-support tools developed for this domain should focus on leveraging existing practices of negotiators, such as information validation and contexualization, instead of automating the negotiation process. Additionally, we recommend prioritizing designs that support training and knowledge sharing over those that directly intervene in the negotiation process, such as real-time negotiation support. 

Finally, our work with the community of practice in negotiation points to the need for ``user development'' -- helping practitioners develop technological fluency so that they are empowered to use AI tools effectively and responsibly. 
The complexity of current LLM technologies poses substantial barriers-to-entry for non-technologically trained communities. We see our work as a call to action for collaborations between technologists and communities of practice that facilitates the transmission of knowledge about AI capabilities and limitations, and supports these communities in developing of informed guidelines for ethical and responsible usage of AI technology. Our current understanding of AI's impact on frontline negotiation is still limited. We hope that future studies will help us clearly separate the responsibilities (and find the complementary potentials) of AI and human in this important application domain. 


\section{Background and Related Work}
\subsection{Challenges in Humanitarian Frontline Negotiation}

International law~\cite{un_universal_1948} requires governments to provide assistance and protection to those living within their jurisdiction. When governments fail to meet this obligation, humanitarian organizations address these violations~\cite{clements_humanitarian_2021}. Within these organizations, humanitarian negotiators work to secure assistance and protection for people in need~\cite{sutton_empathy_2022}. They also engage in face-to-face interactions with stakeholders who could help them reach the people that need help. These stakeholders include but are not limited to the affected communities, donors, and fellow humanitarian organizations~\cite{sutton_empathy_2022, shakun_emotion_2010}. 

Despite the importance of their work, the negotiators often come from a position of weakness, as they need to rely on humanitarian principles or international laws that sometimes mean little to the counterparties. Moreover, they often deal with armed groups while unarmed or negotiate with local governments supported by military forces~\cite{CCHN2019, mancini-griffoli_humanitarian_2004}. 

Due to power asymmetry and ideological differences between the negotiating parties, frontline humanitarian negotiators often struggle to reach principled agreements or accept compromises. Humanitarian negotiations often rule out finding a middle ground, as doing so may involve intolerable concessions or satisfying illegal and immoral interests. Negotiators frequently encounter situations where compromises would lead to outcomes that violate core humanitarian principles. Moreover, when concessions are made, the risk of further compromises to humanitarian goals increases~\cite{CCHN2019}. Constant context analysis, reevaluation and renegotiations are required to navigate these challenging negotiation environments. 

To effectively navigate the dangers and power dynamics faced by humanitarian negotiators, they must employ strategic tools that enable both flexibility and adherence to core humanitarian principles. In high-stakes situations where the negotiating landscape can shift rapidly, robust frameworks are essential for interpreting information and identifying viable paths forward. Organizations like the Frontline Associates, drawing on decades of field experience, have developed frameworks and tools to support negotiators in these tasks. One such tool is the "Island of Agreement," which helps identify areas of shared understanding or common ground between negotiating parties, including agreed-upon facts and converging norms. These shared elements can serve as a foundation for dialogue and trust-building, even amid broader disagreements. (For more frameworks see CCHN Manual~\cite{CCHN2019}).

While these analytical tools are valuable, applying them in real-time negotiations can be challenging. The dynamic and unpredictable nature of frontline negotiations requires negotiators to process shifting positions and facts rapidly, often under intense pressure. In such high-stakes environments, negotiators may struggle to fully utilize these frameworks, as the need for quick decisions and adaptability often takes precedence.

\subsection{Large Language Models and Prompt Engineering}

Current AI research in negotiation largely focus on goal-centric approaches, often overlooking the process (e.g., relationship-building) that leads to successful negotiation outcomes~\cite{zhang2024recommendationsbackwardforwardai}. For example, some works argue that human emotion and rapport-building can hinder negotiation~\cite{cummins_friend_2024}. These goal-centric approaches often involve developing realistic multi-agent simulations that aim to ``surpass human performance'' by negotiating on behalf of humans~\cite{kramar_negotiation_2022, Abdelnabi2024Cooperation, guan2024richelieuselfevolvingllmbasedagents, meta_fundamental_ai_research_diplomacy_team_fair_human-level_2022, Keizer2017EvaluatingPS, Jaidka2024It}. 

In zero-sum, narrowly defined negotiation settings (e.g., contracts negotiation), AI agents are guiding human actions or replacing humans in negotiations altogether.
For example, CICERO, an AI system, has demonstrated ``human-level performance'' in Diplomacy, a strategy game involving negotiation, without human intervention~\cite{meta_fundamental_ai_research_diplomacy_team_fair_human-level_2022}. In practice, companies like Walmart, using Pactum's technology, have automated supplier negotiations, achieving agreements with 64\% of its suppliers, far exceeding their 20\% target~\cite{vanhoek2022walmart}. Similarly, Nibble employs automated chatbots to negotiate discounts for customers on shopping websites~\cite{nibble2021}. LLM chatbots, often viewed as helpful AI ``assistants''~\cite{anthropic2024systemprompts}, typically prioritize outcomes by offering direct solutions and problem-solving advice, bypassing emotional understanding or self-reflection~\cite{chiu2024computationalframeworkbehavioralassessment}.

However, these goal-centric approaches do not align with the realities of frontline negotiations, where emotions, cultural contexts, and irrationalities play crucial roles~\cite{CCHN2019, mancini-griffoli_humanitarian_2004}. In contrast to AI systems focused on outcomes, frontline negotiators must navigate human elements to build trust and secure cooperation~\cite{clements_humanitarian_2021, sutton_empathy_2022, CCHN2019}. For instance, existing non-AI tools in humanitarian negotiation, such as frameworks designed to address emotional dynamics (see Tool 8: Addressing the human elements of the transaction~\cite{CCHN2019}), emphasize the importance of emotions and cultural sensitivities in these complex settings.

Due to the emotional, cultural and other complexities of frontline negotiations, the search space of negotiation strategies becomes far more intricate, making it challenging for current AI systems to replicate human performance~\cite{bbc2021amazon, barrett_emotional_2019, prabhakaran2022culturalincongruenciesartificialintelligence}. This disconnect between AI-driven negotiation trends and the human-centered practices of frontline negotiators points to the need for alternative AI approaches. One such alternative is AI that supports forward reasoning~\cite{lim_diagrammatization_2023, yang_harnessing_2023, zhang2024recommendationsbackwardforwardai}, which focuses on aiding the decision-making process rather than merely generating recommendations. This process-centric AI can enable users to consider their emotional and contextual factors, allowing them to apply their domain expertise independently from the AI's suggestions. By shifting attention from results to the process, such AI systems would be more aligned with the needs of frontline negotiators, assisting them in analyzing contexts and motives in complex, emotionally charged negotiation scenarios. While there isn't currently an AI specifically designed for frontline negotiation, the current trend of negotiation chatbots—which tend to either instruct users or negotiate on their behalf—suggests that future AI developed for frontline negotiation may follow a similar approach.

\section{Formative Study: Understanding the Decision-Support Needs of Frontline Humanitarian Negotiation}
\subsection{Overview}
As frontline humanitarian negotiations grow increasingly complex, practitioners are exploring how ChatGPT can support their work. One member of our team, an experienced frontline negotiator with decades of critical mission experience, worked with us in generating the research questions. Prior to our collaboration, this negotiator and peers in their community have already found ways in which ChatGPT can be helpful to their work, and are equally concerned that these tools can be harmful. Thus, our study focuses on three key aspects: how negotiators are currently using LLMs, the outcomes they hope to achieve, and the concerns they have about applying these tools to their work. Because the negotiator co-author did not want their own views to disproportionately impact the results of the research, they did not participate in data collection or analysis in either of the two studies. 

\subsection{Survey}
We reached out to humanitarian frontline negotiators through the email list of the Frontline Negotiators Network, inviting them to participate in a voluntary survey. The survey aimed to gather information on the participants' negotiation experience, frequency of computer usage for work, educational background, and the extent of AI usage in their professional activities, particularly in relation to frontline negotiation. The questions can be viewed in appendix~\ref{apx:survey}

In total, we collected 30 responses, with 14 negotiators scheduling an interview with the first author. 

\subsection{Semi-structured interviews}
We conducted semi-structured interviews with 14 participants. Prior to conducting our interviews, we made sure each participant provided informed consent, during which we emphasized their right to withdraw from the study at any time if they felt uncomfortable. After completing the interviews, participants received a compensation of US \$30 for their time. Interviews typically lasted 45 to 60 minutes. 

The interviews were structured into three main segments: understanding the participants’ negotiation workflow, discussing their experiences with AI tools like ChatGPT, and exploring potential areas where AI could further supported their negotiation processes. During the second stage where we asked the participants' experience with AI tools, we showed the participants a video of ChatGPT generating the Iceberg CSS text visualization to facilitate future-state ideation based on existing LLM capabilities. The prompts and outputs used can be found in the online supplementary files. 

Detailed interview guidelines are provided in the appendix~\ref{apx:formative}. Following each interview, conversations were transcribed, anonymized, and analyzed to extract key insights.

\subsection{Data analysis}

The first author independently coded 5 interview transcripts using an open coding technique~\cite{burnard_method_1991}. This initial analysis revealed overarching benefits, specific advantages for negotiators, and their primary concerns. Subsequently, three members of the research team convene to finalize a codebook for further analysis. The first author then proceeded to analyze the remaining transcripts, continuously refining the codebook based on new insights until data saturation was reached.

\section{Results - Study 1}
This section summarizes the main findings of our interviews, focusing on the emerging design opportunities and concerns regarding the use of LLMs in humanitarian negotiations. We adopt negotiators' use of term ``AI'' to refer to LLMs.

\subsection{Design Opportunities}
\subsubsection{Summarizing Texts and Context Analysis}
Negotiators often deal with long documents, unstructured texts, and the need to update their preparation in light of new information. Negotiators indicated that steps during context analysis could potentially be assisted by LLMs: \textit{``Providing guidance and support on analysis is where AI can be a game changer. [AI can help finding] what are [different parties'] positions, interest, and needs.''} (P10) In fact, some negotiators have already started using ChatGPT to summarize cases for them. However, prompting still poses challenges. Negotiators want more support on how to prompt LLMs: \textit{``I am not a professional user for ChatGPT [...] if there is any official support [on prompting], I will be very happy.''} (P2)

\subsubsection{Support Understanding Compromises and Associated Risks}\label{sec:ideation}
Effective negotiation involves understanding potential compromises and their associated risks, which negotiators think that LLMs can assist by proposing alternative plans for negotiators to consider, helping them understand a wider range of options and their associated risks. One participant highlighted this potential: \textit{``I just don't want to let my brain be single minded. [...] Maybe AI could be an eye opener for the other way around and proposing secondary solutions and options.''} (P3) Furthermore, LLMs seem to assist in identifying creative compromise solutions that balance the needs of all parties involved. As one negotiator suggested, LLMs could aid in \textit{``Finding creative solutions and new positions that can satisfy all parties' needs.''} (P10)

In the context of risk assessment, negotiators believe LLMs can help identify information gaps that could lead to unforeseen risks in compromises. One participant noted, \textit{``You could cross check [with AI] what you’ve got. This is what I’ve got on this counterpart. Is there anything out there I’ve missed. That would be very useful.''} (P12)

\subsubsection{Support knowledge sharing}
Knowledge sharing between negotiators involves new negotiator training and sharing experiences of past cases. P14 shared in their decades of negotiation experience, they have \textit{``lost 22 colleagues''} to violence and emphasized, \textit{``With so many conflicts happening, negotiators don’t have the luxury of 10 or 20 years to learn. I hope AI can help new negotiators build skills without facing hostile environments.''}

Another highlighted the loss of institutional memory, noting, \textit{``We lost incredible knowledge about negotiations from 20 years ago, even with the same people and challenges.''} They stressed capturing strategies and common ground, not just outcomes: \textit{``What I care about is the strategy, the common ground. A dataset that proposes past cases from various organizations would be invaluable.''} However, confidentiality often limits shared learning: \textit{``Maybe AI can be trusted enough to decontextualize these cases.''} (P10)

Sharing past experiences while preserving confidentiality could inspire new strategies and help predict opponents' moves: \textit{``Having access to information on past decisions or insights into a counterparty's foreign policy is great. Knowing how a country positioned itself on global forums allows you to place them on a spectrum in terms of this position.''} (P11) 


\subsection{Concerns Over LLM in Negotiation}
\subsubsection{Confidentiality}
The information that the negotiators have to handle often requires consideration over confidentiality. For that reason, during negotiation training, real cases for education purposes are anonymized. Negotiators proposed that common confidentiality practices should be applied to usage of LLMs too. \textit{``[Confidentiality] is a very big concern, because when it comes to data, it is very sensitive in these conflict areas. We need to really protect those who are coming out to share those information.''} (P5)

Moreover, negotiators felt that they were not equipped with adequate understanding of LLM, particularly concerning privacy and the appropriate usage of sensitive information. \textit{``My organization has already put in place some kind of restrictions on the way we can use AI but we are still in the beginning of steps of understanding the capabilities and also the potential consequences, and so on. The guidance we have received is still very vague and very broad.''} (P8)

Although there's hesitation over confidentiality, negotiators who have been using ChatGPT in their workflow gradually felt more comfortable after using the tool for a while. \textit{``[Confidentiality] is actually what I was very worried about at the start. [...] But with ease of use, and seeing how useful it was, I then started to feel much more comfortable. I still don't put high security things or security protocols or confidential documents into it. But I feel much more at ease to put transcripts in and ask for summaries of documents.''} (P4)

\subsubsection{Western Bias in LLMs}
Negotiators raised concerns about the Western bias embedded in LLMs like ChatGPT, which limits their effectiveness in diverse cultural contexts. One negotiator noted that ChatGPT largely reflects the perspectives of its creators, who possess a particular level of English proficiency, education, and cultural understanding. This bias is problematic since the tool is meant for field teams globally, not just its developers. P12, a European negotiator, emphasized: \textit{``You know, this is meant to be useful for negotiators [in non-Western countries], not for me to be honest. You need a lot, a lot of feedback, but not from us [...] You really need to go to Nigeria, to Myanmar, and talk to the field teams and get their inputs on all these.''} 

Bias in LLMs also affects users from non-Western contexts who are less familiar with Western-developed technologies. For example, locally recruited negotiators may struggle with certain interface designs. As P10 explained, \textit{``Our negotiators are usually locally recruited staff with very basic skills for interacting with AI, and not the best English skills, for instance. They may also lack the ability to create proper reports to feed into the AI.''} 

The bias within LLMs not only affects users from non-Western backgrounds but also impacts Western users, as negotiations often span different cultural contexts. \textit{``You really want to avoid that bias. I don’t want to just draw information from a Western point of view when I can draw information from the context or country where the negotiations happen.''} (P12)

\subsubsection{Public opinion/ mandators' opinions on AI}
Negotiators often operate within the red lines and limits set by their mandators and consider the well-being of local communities. A mandator, such as the head of a humanitarian organization or a country, sets the strategic objectives and boundaries for negotiators, for example, by providing guidelines on terms like ensuring aid worker safety and maintaining neutrality during conflict zone negotiations. The opinions of both mandators and the public significantly influence whether and how negotiators would think about using LLMs. One negotiator voiced concerns over using LLMs without \textit{``taking the time to really explain and understand what the perception of local communities is [on AI].''} Similarly, mandators such as governments' opinion on LLMs can impact the usage of LLMs in negotiation. \textit{``[...] I'm not sure governments would be very happy or trustful to know that we're using AI in order to negotiate for them.''}(P10) 

The doubts may come from lack of understanding of these tools. Negotiators think that more public understanding about the limitations and benefits of these tools will facilitate the conversation of whether to integrate LLMs into negotiators' workflow. \textit{``I feel if there were loads of people who understood it a lot better, they’d feel more comfortable using LLMs in negotiation, because we all have these myths about what this stands for. [...] There should be a whole education on the reality of these tools and what they can offer. I don’t think people are very well informed.''} (P12)

\subsubsection{Reliance and trust}
Similar to other AI tools, some negotiators are concerned over LLMs' reliability in negotiation tasks. \textit{``This is talking about my experience, but you cannot depend on it entirely 100\%, because it's still a tool.''} (P2) Often times confusing or inaccurate answers force the users to consider the reasons behind LLMs' generation of such answers. For example, P3 asked ChatGPT about what set of milestones they should set for their current project. \textit{``AI might come up with 2 different set of milestones. So my question is, what's the reason of these different results? To which extent should I rely on it?''} (P3) Another negotiator mentioned a similar point, citing that without knowing the source of why LLM generates certain answers, they are not able to rely on these systems. \textit{``What information is the AI basing this on? And why is it choosing [...] those specific documents to answer your question? I don't know. I'd be curious to know what the parameters that the AI based on''} (P12)

There are instances where incorrect LLM results are acceptable, especially under time constraints when \textit{``it’s better than nothing''} (P2). Negotiators also see LLM outputs as sources of inspiration, even when inaccurate. One noted, \textit{``Even if AI is not right, we can tell. It’s okay because it challenges our ideas''} (P4), comparing AI to an \textit{``aspiring partner''} that stimulates thought. Another mentioned using AI to review transcripts and ask, \textit{``What are questions I didn't ask? What should I have asked for more details?''} helping refine their ideas.

\subsubsection{Practical Limitations of AI in Negotiation}
The effectiveness of AI tools like ChatGPT in negotiation relies heavily on the quality and structure of the input. Negotiators noted that significant human effort is required to prepare information, such as converting meeting notes, emails, or phone call records into structured formats for AI to process. Moreover, negotiators often lack structured notes, particularly when handling sensitive or confidential information that isn't formally documented. As P11 explained, \textit{``If this [case file put into ChatGPT] originates from existing notes or meetings ... sometimes we don’t have those notes. [...] After a phone call, I might jot down two points, but I wouldn’t have a full script or source for this [to input into ChatGPT].''} However, they also thought that AI might encourage negotiators to be more organized: \textit{``It may enable a negotiator who is all over the place to be more structured. And so it can be a positive thing.''}

\subsubsection{Overreliance}
Negotiators are concerned that over-automating the negotiation process might affect one's ability to conduct negotiations effectively. One negotiator expressed this worry using an analogy:
\textit{``I think that all these tools will shrink our brains because we will start relying on them. Like, it's the same as with GPS. [...] We stopped noticing the roads we take before the destination. And suddenly, you are there? How? You don't know, because machine was leading you.''} (P3)
Another negotiator emphasized the importance of manual analysis:
\textit{``I think it's important to do your own analysis. I think that information and intelligence comes from doing proper analysis and getting yourself to a certain spot. [...] My worry is lack of engagement in the analysis.''} (P4)
\section{Understanding LLM in Negotiation with a Probe}
\subsection{Overview}

Study 1 revealed negotiators' initial hopes and concerns about using ChatGPT in negotiations. It also revealed that some negotiators have already begun to experiment with ChatGPT. To further study the practical challenges, including prompt engineering, we designed a second study focusing on negotiators' experiences when actually using LLMs in negotiation contexts. 

We explored two ways of using LLMs. One---the current baseline use---was to give participants direct access to the unmodified ChatGPT interface. The second was a custom-designed probe interface. We developed the probe with Process-Oriented Support frameworks. Unlike goal-oriented AI, this approach supports users throughout the decision-making process rather than focusing solely on outcomes. It also obviates the need for the negotiator to develop their own prompt engineering skills.

The probe interface, compared to ChatGPT, aims to:

\begin{itemize}
    \item Reduce prompt engineering's impact on usability perception;
    \item Investigate current negotiators' practices through conduct a realistic case to identify aspects relevant to LLM applications, enabling a comparison between negotiation practices and what existing goal-oriented AI support systems try to assist;
    \item Broaden negotiators' perspective on LLM capabilities so that they can envision future LLM impacts in negotiation.
\end{itemize}

\subsection{Probe Design Overview}
\begin{figure}[h!]
    \centering
    \begin{subfigure}[b]{0.45\textwidth}
        \centering
        \includegraphics[width=\textwidth]{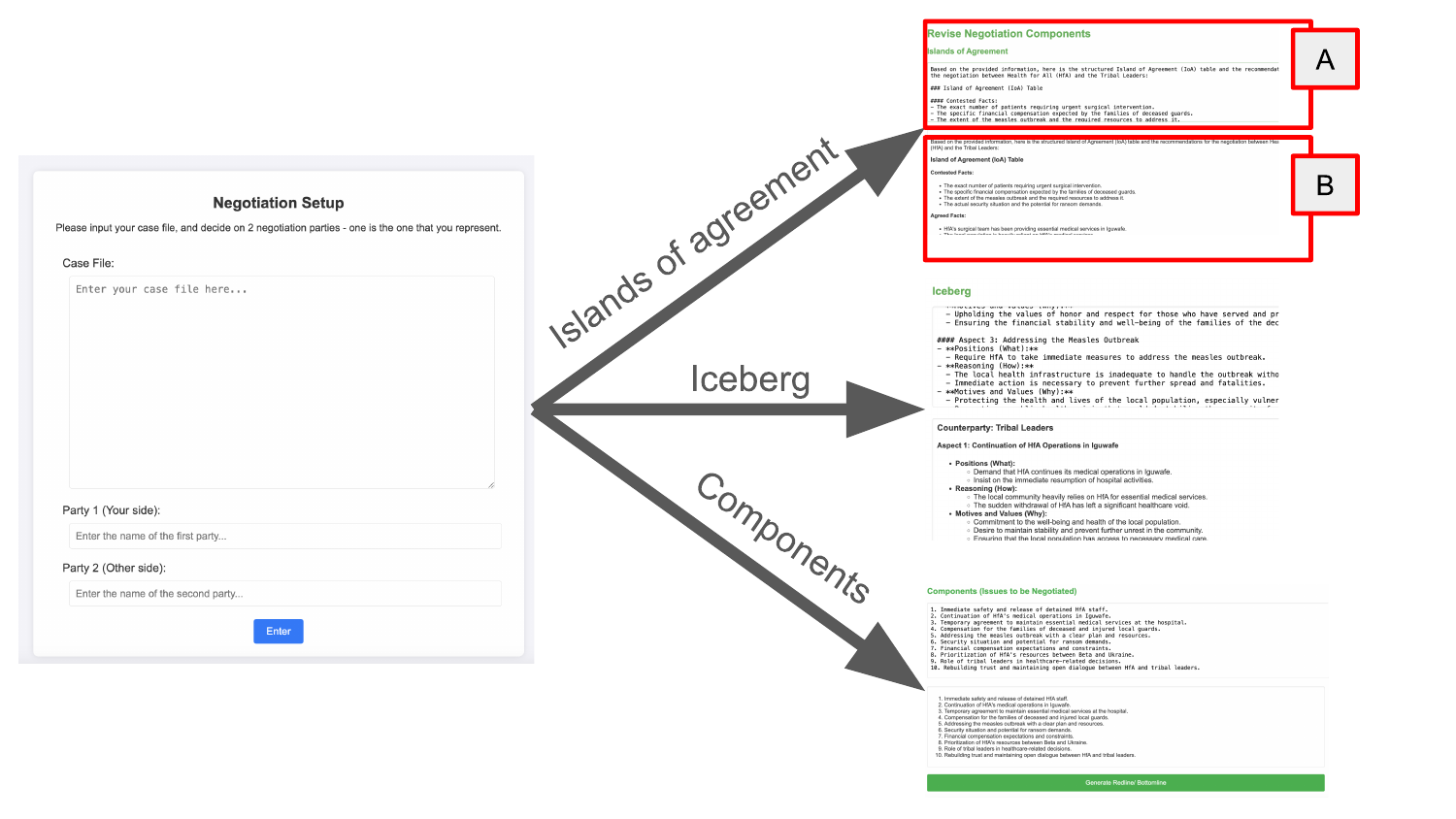}
        \caption{At the left, the user can input the case file, and the negotiation parties. Then the probe will generate Islands of Agreement, Iceberg CSS, and Components. The user can choose to inspect the generated content in B in markdown format, and modify in a text box like A. }
        \label{fig:svg1}
        \Description{This subfigure shows the initial input interface where users enter case data and negotiation participants. It assists in generating visual representations to aid in understanding the negotiation context.}
    \end{subfigure}
    \hfill
    \begin{subfigure}[b]{0.45\textwidth}
        \centering
        \includegraphics[width=\textwidth]{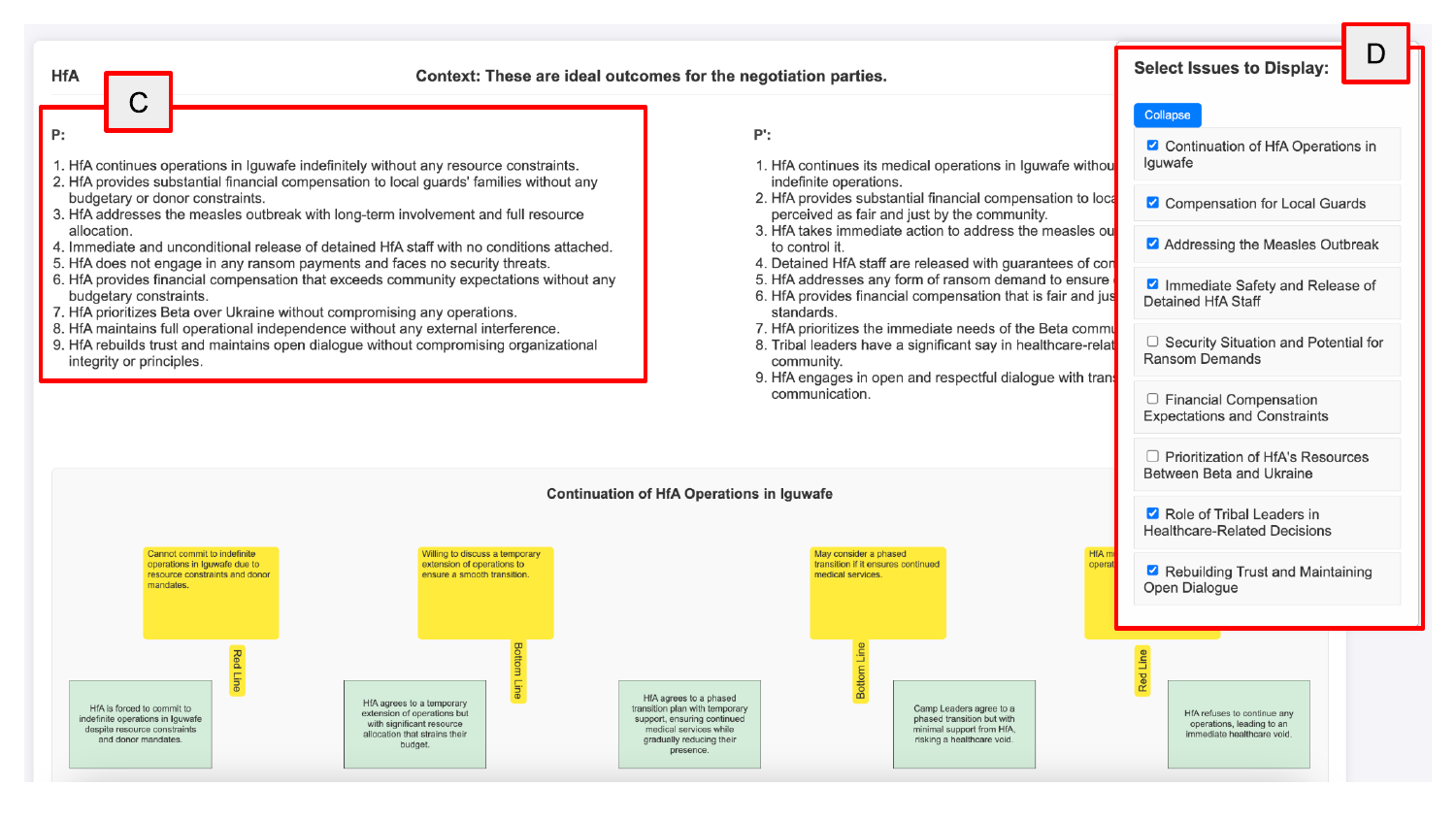}
        \caption{This figure shows the redline and bottomline (RL/BL) tool. Panel C represents the ideal outcomes for the party represented by the user. These outcomes serve as a starting point for the possible compromises in the RL/BL tool. Panel D lists all the issues that will be negotiated. Users can choose or reorder issues from panel D, which will be reflected in figure~\ref{fig:svg3}. }
        \label{fig:svg2}
        \Description{This subfigure visualizes the RL/BL tool, outlining possible negotiation compromises and issues. Users can see their party's ideal outcomes and reorder issues for negotiation.}
    \end{subfigure}
    
    \vspace{1cm} 
    
    \begin{subfigure}[b]{0.6\textwidth}
        \centering
        \includegraphics[width=\textwidth]{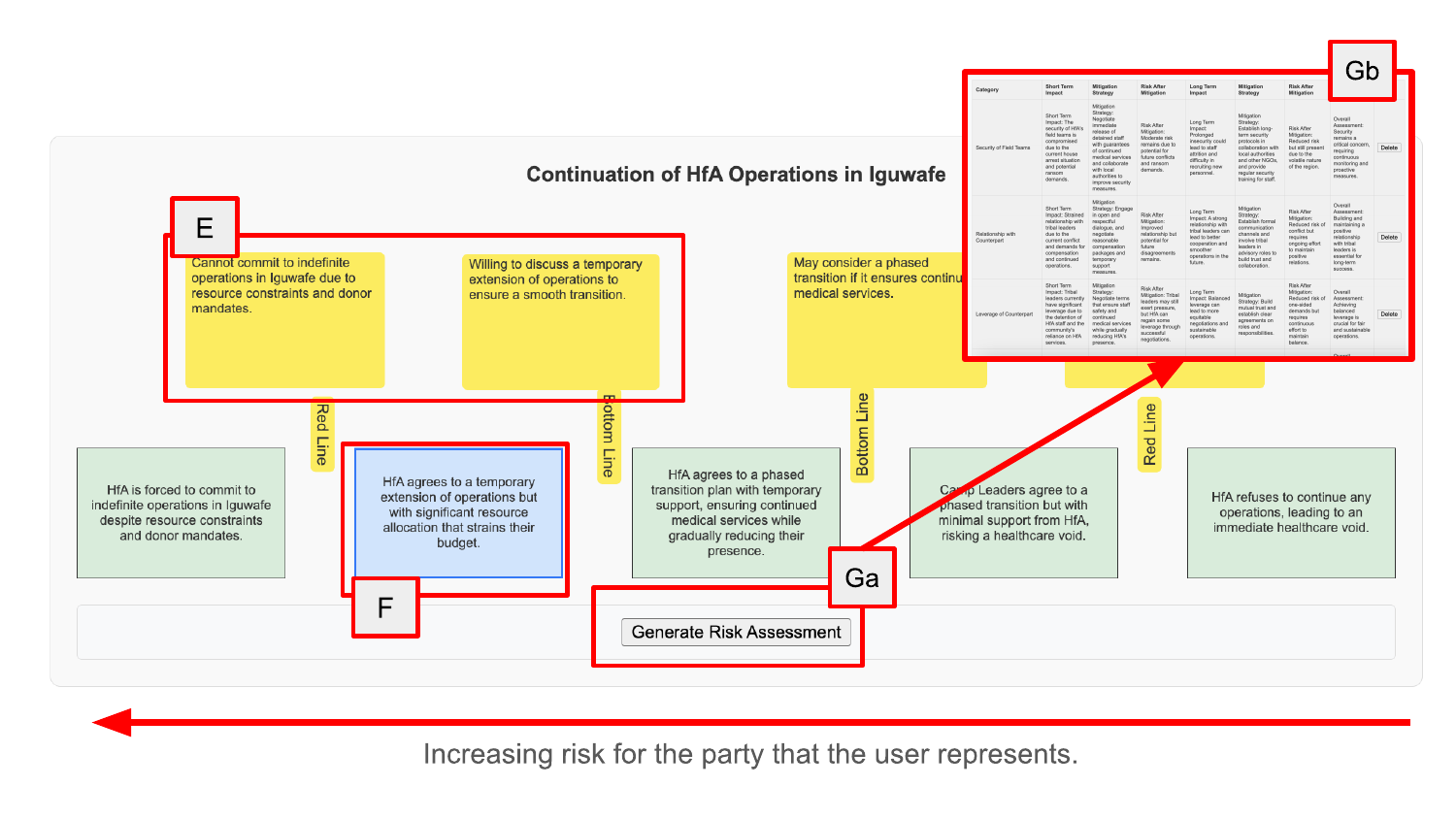}
        \caption{Panel E shows the redline and bottomline of the negotiation party that the user represents. The two boxes on the right show the redline and the bottom line of the counterparty. Panel F shows one of the scenarios or outcomes that the negotiator decides on. The outcomes are listed such that, from the right to the left, there is an increasing amount of risk of negotiation failure as the outcome crosses the bottomline and the redline of the party that the user represents. After selecting the scenario, the user can generate a risk assessment for this selection through button Ga, and show a risk matrix in Gb. }
        \label{fig:svg3}
        \Description{This subfigure illustrates the redline and bottomline of both parties in the negotiation. It helps negotiators visualize the increasing risks as they approach or cross these lines, with the option to generate a risk assessment.}
    \end{subfigure}

    \caption{Overview of the probe interface.}
    \label{fig:svgset}
    \Description{The figure illustrates different sections of the probe interface used in humanitarian frontline negotiations. In panel (a), users input the case file and parties, while panels (b) and (c) demonstrate tools for redline, bottomline, and risk assessments that guide the negotiation process.}
\end{figure}

We designed a probe interface based on the design opportunities in study 1 for a bilateral humanitarian negotiation. We used OpenAI's GPT-4o API~\cite{OpenAI2024GPT4} as the underlying LLM. The prompts used in this interface can be found in online supplementary materials. We emphasized the designs for tools that help the negotiator summarize contexts, ideation, and navigate risks and compromises. We describe these design features in this section. An overview of the probe can be seen in figure~\ref{fig:svgset}.

\textbf{Contexualization with frontline negotiation frameworks.}
The probe begins by collecting the case file input by the negotiator. Users then identify the negotiating parties involved. The system summarizes the case file into key elements: the Island of Agreement (IoA), Iceberg CSS, and the negotiation components to be addressed. The prompts for generating the IoA, Iceberg, and Components were created using the ChainForge interface~\cite{Arawjo2024Chainforge}, with input from a co-author with over 30 years of frontline negotiation experience. These prompts were designed and evaluated to ensure the most suitable outputs. The IoA and Iceberg CSS tools summarize the positions of the parties, while the Components represent individual negotiation elements critical for success. Once the text summaries are generated, users can validate and modify the content in the provided text boxes.

\textbf{Find Zone of Possible Agreements with Bottomline and Redline}
A redline represents a compromise that crosses an unacceptable level of risk or violates core principles, such as neutrality or humanitarian mandates. A bottom line, by contrast, marks the point at which negotiators are willing to walk away, where the benefits no longer outweigh the risks. The interface presents these limits on spectrums for each negotiation component. It categorizes compromises based on whether they violate the user party’s redline, bottom line, or fall within the zone of possible agreement, as well as the counterparty’s limits. The interface visualizes these red and bottom lines on spectrums, showing scenarios where 1) the user’s redline is violated, 2) the bottom line is violated, 3) neither line is violated (zone of possible agreement), 4) the counterparty’s bottom line is violated, and 5) their redline is violated. Each step to the left indicates greater risks for the negotiator’s party, with each compromise pushing closer to an unfavorable outcome.

\textbf{Generate Risk Assessment.}
The user can click on any of the scenario boxes to indicate that they are considering picking the outcome for that specific component of negotiation. The user can choose to let the interface generate a table of risk assessment, indicating the short term risks and long term risks, mitigation strategies, and risks after mitigation. These assessments are generated by categories: Security of Field Teams, Relationship with Counterpart, Leverage of Counterpart, Impact on other organizations/ actors, Beneficiaries/ Communities, and Reputation. These standards are the common practice from a negotiator we interviewed in study 1. The users can modify these columns directly. 

\subsection{Methods}
\subsubsection{Participants}
We recruited 18 negotiators different from study 1 from the same survey results. We provided each participant with \$30 Amazon gift card to thank them for their time. 

\subsubsection{Study Design}

This was qualitative a within-subjects study that followed the baseline-intervention design (similar to, e.g.,~\cite{trinh2014pitchperfect}) with the order of conditions fixed such that all participants interacted with the ChatGPT condition first followed by the probe interface. The baseline-intervention design is appropriate when the baseline condition is similar to a practice that participants are already familiar with and where the intervention condition may have lasting spillover effects on how people approach the task (our probe interface does that by explicitly focusing on specific parts of the preparation process).

Each study session lasted approximately 60 minutes or over, and was conducted remotely via Zoom. The study consisted of three parts:

\textit{Part 1 (20 minutes):} Participants were given one of two anonymized real-world case files. These case files can be found on the online supplementary material. They were asked to use ChatGPT while sharing their screen over Zoom to develop a negotiation plan from a humanitarian perspective. After 20 minutes or when they felt ready, participants drafted a preliminary negotiation script categorized into three tiers:
\begin{itemize}
    \item Issues and terms that are low-cost and high-benefit for both parties, aimed at building relationships and setting a positive tone.
    \item Issues and terms that involve some cost and benefit for both parties, intended to establish a fair basis for distribution.
    \item Complex issues and terms that are high-cost and central to the conflict, potentially contentious and best addressed later to avoid confrontation.
\end{itemize}
These three tiers were based on the CCHN manual, which recommends categorizing issues in this way prior to negotiation.

\textit{Part 2 (20 minutes):} Participants were then provided with the other case file, which they had not yet seen. They were given access to the probe interface and again asked to share their screen over Zoom. The task was to perform the same analysis using the probe interface. Participants were encouraged to think aloud and share their immediate thoughts during the first two parts of the study. The order of using ChatGPT and the probe interface was not varied, as starting with the probe might have influenced participants to approach the analysis differently (e.g., by breaking the task into smaller steps) than they would when using ChatGPT.

\textit{Part 3 (20 minutes):} In the final part, participants were interviewed about their experiences using ChatGPT and the probe for case analysis, the added value of each system, any issues encountered, suggested improvements, and other AI features they would like to see. A complete set of interview guidelines can be found in Appendix~\ref{apx:probe}.

\subsection{Data Analysis}
As in Study 1, all sessions were audio recorded and transcribed. To analyze the data, we employed thematic analysis with open coding. The first author initially coded the transcripts and identified a preliminary set of themes. Three members of the research team then convened to review the themes and associated data segments, resolve any discrepancies, and refine the theme definitions. The first author subsequently re-coded the data using the finalized themes, and the research team validated the final set of themes during a follow-up meeting.

\section{Results}

\subsection{Common Practices of Negotiations}
This section will discuss the common practices related to frontline negotiation that emerged from the interviews.

\subsubsection{Frontline negotiation is complex and context-dependent.}
Frontline negotiators think that humanitarian negotiation differs from other forms of negotiation, such as business or transactional negotiations, due to its complex context. As P5 explains: \textit{``In the humanitarian world, there is such a myriad of breadth of complexity of environments, social environments that we operate in. Business, well, of course, there are different approaches, but primarily it's very transactional in terms of a negotiation.''} For example, P2 explained how NPOs in Japan are perceived differently from US and Europe: \textit{``Nonprofits are not trusted in Japan. So if ChatGPT said something along lines of `you should seek public support for your project'. All of us in the room would go. `That's a valid answer, maybe for United States, or maybe Europe. But for Japan. based on our past experience, our own understanding of where we stand within society, that would not be a valid strategy.' ''} This complexity requires practitioners to continuously adjust and adapt their negotiations styles in accordance with changing situations (P6).

\subsubsection{Frontline negotiation requires empathy and flexibility.} Being Empathetic and flexible are important qualities for reaching an agreement in negotiation. Yet,\textit{``the human ability to be flexible, to adapt and react [...], and understand the human element and the human dimension of a negotiation''} are missing in current AI tools. (P3) This human touch extends beyond mere strategy to building trust and relationships, sometimes requiring negotiators to patiently invest time and effort in creating a foundation of trust. P5 illustrated:
\textit{``I've sat in various operations where I had to drink tea for weeks on end with a counterpart and not get to the nitty-gritty until they were ready to do that because they felt that they could trust you.''} In other cases, it simply requires the negotiators to act quickly in respond to rapidly changing dynamics: \textit{``Things might get tense. You know. I'm thinking of a negotiation in my case where suddenly gunshots are fired during the negotiation. That changes the dynamic very quickly. Because now I'm really not in favor of the local guards being in charge of security because they're firing guns in the air.''}

\subsubsection{Verifying information in negotiation is iterative:} Negotiation is an interactive process involving continuous information exchange and analysis between negotiators, counterparties, and stakeholders. Its dynamic nature requires ongoing updates as new information emerges and gaps are identified. P1 illustrated, \textit{``Negotiation is not a give, put money, get answered process [like a vending machine]. It's an interactive process, so it means that we collect information, we work on it, and then we analyze it.'' } 

The iterative nature of negotiation affects how information is verified. Negotiators reported verifying information at different levels, from individual checks to group consensus, ensuring accuracy and preventing complications. P3 described their verification process: \textit{``Every time there's been a doubt we verified it. As far as I know, nothing that has been produced as part of my job has led to, you know, mistakes, complications, or diplomatic tensions, because we got something wrong. If we saw something strange, we verified it, and then we kicked it in. If it was correct and we moved it. And we couldn't move without [verification]. ''} (P3) Achieving consensus iteratively can be challenging, especially when cultural factors and individual skepticism complicate the process. P2 highlighted this difficulty, particularly in contexts where unanimous agreement is valued: \textit{``I'm in Japan, where consensus building is a huge issue, that you could have 10 people in a room, and if 9 people agree, that's still not consensus. So, in other words, if there's only one person who's providing an analysis, there could be a sense of `is that really true?'''} (P2)

\subsubsection{Negotiators believe that their work practices can reduce the impacts of AI biases and mistakes on outcomes:} 
Negotiators believe that biases are unavoidable, and managing them is essential in the negotiation process. As P4 noted, impartiality is often an \textit{``ideal perspective''}: \textit{``Impartiality is idealistic—nothing is truly neutral. We are never independent, never neutral, never impartial. Humans compromise on principles.''}

Negotiators conceptualize mistakes as a natural part of the negotiation process, often arising from the differing perspectives of negotiators and their counterparties. These discrepancies are not only tolerated but viewed as opportunities for deeper discussion. Under this system, negotiators believe that inaccuracies, whether human or AI-generated, create room to explore the reasons behind disagreements and the factors contributing to varying viewpoints. When asked if AI-generated factual inaccuracies would negatively impact negotiations, P2 responded, \textit{``No, not at all. [...] I go in with my own assumptions and biases. [...] If I'm aware of those, I can adjust when speaking to the other party. [...] Disputes over facts don't bother me as much. [...] What would concern me is if we both believed we agreed on a fact, but in reality, we didn't.''}

However, mistakes from AI can introduce more ambiguities than those arising from human biases or errors, but negotiators believe that their existing practices can mitigate the impacts of AI errors. As P7 noted, \textit{``I’m generally more hesitant to take things that are generated by an AI and directly apply them simply because of the margin of error and the implications that those errors can have in direct human interactions.''} Despite the AI's limitations, negotiators have developed their own strategies for verification to overcome some of the inherent flaws. P1 explained, \textit{``When I use ChatGPT to analyze something, I always do another layer of triangulation, human intelligence, talk with some other people to verify if this thing, which I thought or which we thought, is okay to go forward.''} P6 provided a similar approach: \textit{``It gave me a direction, but I would need to verify on the ground ... I’d probably find the people that’s closest to me and run it by them to find out what their perspective is.''} P1 stressed the importance of using information from trusted sources, stating, \textit{``If I use ChatGPT, I would instruct it to reference only ICRC-approved materials. I’d only trust the output if it aligns with our official strategy and tools.''} 

Finally, negotiators highlight that evaluating the broader context and relational dynamics in negotiations is often more critical than simply obtaining the correct information; successful outcomes depend not just on the factual accuracy of their information but also on the depth of relationships and trust they build with their counterparts. P3 pointed out, \textit{``It’s important to meet with local stakeholders to build rapport and understanding. [...] Having that kind of information also tells me the impact of going with a certain line could have consequences on the relational aspect of the negotiation.''} 

\subsubsection{Negotiators Feel the Need to Take Full Responsibility When Using AI tools} 

Negotiators stressed the importance of maintaining full responsibility when using AI tools like ChatGPT, unlike situations with human counterparts where institutional power may influence decisions. P2 explained, \textit{``ChatGPT is just a tool. If I choose to trust it and things go wrong, the responsibility is still mine.''} P4 echoed this, stating, \textit{``The responsibility always lies with the negotiator. We can’t blame the tool if something goes wrong.''} With human input, responsibility may shift due to institutional power dynamics or professional networks. In comparison, taking positions from a human can differ because a person may hold institutional power (such as \textit{``a manager who has institutional demand power''} (P1)) or be part of the negotiator’s professional network (for example, \textit{``a colleague who is also in the negotiation''} (P1)), so that negotiators are able to transfer some of the responsibilities to them.

\subsection{Benefits of LLM-powered applications}

\subsubsection{LLM-Powered Applications Save Time}  

Our results showed that both ChatGPT and the interface can significantly reduce the time needed for negotiators to process and analyze information. P1 highlighted ChatGPT's ability to condense years’ worth of meeting minutes: \textit{``It saves me time from reading it all. It helps me condense information from this negotiation.''} Similarly, the interface streamlines tasks like risk assessments and planning, which would otherwise be time-consuming. P1 noted, \textit{``You just eliminated hours of meetings,''} while P5 emphasized, \textit{``It's extremely swift. This would normally take my team considerable time.''}

However, negotiators acknowledged that while these tools enhance productivity, they don't address the core aspects of negotiation. P4 admitted that using ChatGPT felt like \textit{``cheating''} for tasks negotiators should already handle. Similarly, P1 noted that the efficiency gained doesn't replace human expertise and judgment: \textit{``The Interface saves time, but if it didn't exist, I'd still need to manage these scenarios on my own. I can also ask a manager, I can ask a more experienced colleague, I can relate.''} P7 echoed this sentiment, noting that while the tools are helpful, they can only supplement the existing systems of knowledge and experience that negotiators depend on: \textit{``It can only be in addition to already existing systems, knowledge, experience.''}

Negotiators also stressed that these tools are most useful during the preparation stage, but their benefits diminish during the actual negotiation. P1 highlighted the importance of applying human skills to use the information generated by the Interface: \textit{``If someone has this strategy [generated by the Interface], but as cold as it could be, lacks negotiation skills, this is just a document and this is just information. But how to use information is important with the complete package of the negotiation.''}

\subsection{Limitations of LLM-powered applications}
\subsubsection{Automating the Human Elements Away}

Negotiators expressed concerns that current AI tools fail to capture key aspects of human interaction. As P5 explained, \textit{``It's often the soft stuff [...] like body language and tone of voice''} that plays a significant role during negotiations. This limitation means essential non-verbal cues, important for understanding and influencing counterparties, are lost. P4 echoed this, warning that over-reliance on AI could \textit{``remove some of the human components from the whole negotiation.''}

Beyond non-verbal elements, there is a broader concern that AI's potential widespread use in negotiations could reduce the authenticity and depth of human interaction. P5 warned that relying on AI might result in \textit{``less genuine interactions''} as negotiators miss subtle cues like facial expressions, crucial for understanding motivations. This could lead to what some describe as a \textit{``sterilization of the negotiation process.''} P5 further explained, \textit{``If we all use this, [...] we lose some of the thought process,''} and potentially negotiation parties would \textit{``find fewer options, as negotiators may not come up with the most creative ways to resolve a deadlock''}(P16). 

However, these limitations highlight the distinct roles AI tools and human negotiators play. AI can support negotiations by providing context and aiding in strategy development, but it cannot replace the human elements that build trust and relationships. P4 noted that \textit{``most of the humanitarian negotiations normally take into consideration that we have to have proximity with the person, we have to have empathy, those kind of things. And then in this kind of tool, that might not be the case...''}  Negotiators believed that while AI can assist in many ways, it cannot replicate the empathy and interpersonal connections that are often the cornerstone of successful negotiations. \textit{``[Human emotions and alike] wouldn't be something that you would be able to necessarily input in there...  But it's not designed to do that... It is not supposed to capture the diplomatic aspect of the negotiation or the human aspect of the negotiation.''} (P3)

\subsubsection{Limitations of Resource Intensive Technology}
LLM applications usually requires fast Internet connections or locally deployed models with powerful GPU. As humanitarian negotiations often happen in resource scarce areas, it is difficult for users. \textit{``Most of the people who are involved in negotiation don't have high speed Internet connection in the field. If the people are under pressure, because we work in high pressure environment normally, and then this small delays in the timing [due to Internet connection] might make them using it less.''} (P4)

\subsubsection{``Overconfidence'' of LLM tools}
ChatGPT always appears to be confident, which makes it hard for negotiators to identify uncertainties: \textit{`` Well, it will answer with great amount of certainty. And so that's a caution for making sure that you also feel confident about what it's the information that it's giving back to you so that you, you would need to know enough about the situation to go `ehhh that doesn't sound quite right'.''}  (P12)

What makes ChatGPT harder to trust is its inability to proactively identify or inquire about missing information. Unlike human negotiators who ask clarifying questions, LLMs provide answers based solely on the given input without flagging gaps. P1 pointed out that if something is not mentioned in the document, \textit{``the generator will not mention it as a suggestion.''} P1 also shared how they got the LLM to ask questions by breaking tasks into smaller parts and providing explicit instructions, but noted that this requires extra effort.

\subsection{Comparing ChatGPT and the Interface}
\subsubsection{Chat interface lacks structure or guidance}
Negotiators, even those with considerable experience using ChatGPT, found that prompting is difficult. Over half of the participants (n=10) copied and pasted entire case files into the chat window, hoping to get results in a single round. Others developed multi-step prompting strategies. For instance, P1 \textit{``separated the context from the case''} and generated a background summary before moving on to detailed tasks. Despite this, P1 still found it difficult and asked ChatGPT for help in refining prompts: \textit{``I asked ChatGPT to analyze the instructions and suggest improvements… it would tell me, `this is confusing, better to phrase it this way.' ''}

These challenges with ChatGPT can especially impact less experienced negotiators. Participants suggested that novices might be more likely to over-rely on AI, not question its outputs, or fully grasp its limitations. As P6 explained, \textit{``If I didn’t have a strong background in due diligence, I might be overly confident in my assessment. I’m more concerned about novice negotiators using ChatGPT and assuming it’s correct. Experienced humanitarian negotiators have enough buffers around them to avoid major mistakes, but novices could misstep.''}

On the contrary, our probe interface seems to help level the playing field between negotiators of varying experience levels. P5 pointed out that if only the experts can generate the necessary frameworks like Islands of Agreement then \textit{``the success of a negotiation would only depend on the expert negotiators' [...] With this tool, teams can work better, faster, in shorter time, and they don't need the huge experience in negotiation.''} and the probe interface can help less experienced negotiators generate visualizations and analyses that typically require expert knowledge. 

However, expert negotiators argue that they don't require many guardrails but instead need flexibility from AI tools to tailor outputs to their specific needs and adjust strategies dynamically. When asked how the interface could be improved, P11 suggested: \textit{``...allowing more direct interaction with the interface could be useful. For example, if I can make some changes here directly, even without altering the original documents, that would be helpful.''}

\subsubsection{Showing Negotiation Process rather than Recommendations}
Negotiators emphasize the importance of having options and flexibility rather than receiving a fixed recommendation from ChatGPT. Negotiators want ChatGPT to engage in a collaborative brainstorming process with them, rather than dictating actions or providing specific recommendations that limit adaptability to varying contexts. \textit{``I’d rather the AI poses those questions, what compromises are you comfortable with? And here are the different options, and then it’s up to you to choose which one... when it says prioritize and avoid, that's up to me to decide. That's not up to the AI to decide. In certain situations, I might take more risk because there's more lives at stake or the situation is by far worse, but sometimes I might take less risk if the situation is not as bad or if there's a possibility that there might be another solution that might come up.''} (P8)

Negotiators felt that the probe interface, by not offering a fixed strategy, enhanced collaboration by allowing them to explore options and develop context-specific strategies.  P3 emphasized, \textit{``[The interface]'s set up to help you organize your ideas, give you a comprehensive view of everything, and not to give you every solution or guarantee you a solution. I think it’s excellent.''} Others appreciated that this approach led to less biased decisions.  \textit{``What your program did was it actually kind of removed my initial sensory inferences that may have clouded my assessment.''} (P6) This observation echoed with other negotiators who prefered that AI tools only give the users options rather than recommendations: \textit{``I rather these tools give us just different options with different risks associated with it, and then just say, okay, user, it’s up to you what you want to do, what choice you want to make.''} (P8). Additionally, to the negotiators, our interface that showed the process aided in the ideation process, helping teams brainstorm and outline potential solutions. As P2 explained, \textit{``The benefit of this model is that a lot of those things are in my head, more at the subconscious level. But to act on them and make them useful, I’ve got to bring them to the conscious level. When I see it here, I can say, `Oh, yeah, I see how that aligns with what I’m thinking.' ''}

\subsubsection{The Interface Encourages Collaborative Communication}

In contrast to ChatGPT, participants expected that the probe interface facilitates teamwork by making it easier to present, discuss, and revise cases collectively. By generating structured outputs, the interface could potentially simplify projecting information to the team and engaging in productive discussions. As P1 noted, \textit{``It's easier, not just for the individual, but for teamwork. When I prepare a case and put it on the Interface, it’s easier for whoever uses it to project it to the team and then have a discussion.''} Echoed P5: \textit{``it may help us not just with the negotiation preparation itself, but it's going to help us considerably be able to communicate in a very digestible manner from the different levels of people involved in a complex negotiation.''}

This interface can leverage negotiators' existing information validation practices to cross-check LLM-generated outputs. P2 described how they would use the interface to validate information with their peers: \textit{``The added value is, if I have a negotiation team, and collectively, we write up the case study, upload it to the interface, and it gives us islands of agreement, we can see where we agree or disagree on facts and norms. It's helpful, especially if not everyone has authored the case study. Running it through the interface provides a simplified way of looking at the data.''} P2 further explained how the interface could help categorize information by relevance and accuracy: \textit{``I’d categorize information into three groups: one, confirming what I know; two, definitely incorrect or incomplete; and three, things I need to verify. Even the ‘let me check that out’ category adds value, as it becomes something I need to confirm during the negotiation and assess its importance.''}

The participants thought that the interface not only facilitated collaboration but also helped negotiators clarify boundaries and align with stakeholders. P11 discusses how the interface could potentially help in planning and strategy development, allowing negotiators to clarify what can and cannot be done, which is important for aligning with the mandate and gaining consensus among stakeholders. \textit{``I would use that in the planning, specifically in planning to elaborate the strategy and the planning of the different steps, and also for sharing with the mandate subject. [...] So you will clarify the mandate subject, what I can do or not do, and to clarify the mandate more clearly also.''} (P11) The negotiator mentioned that they would use the Interface for strategic planning, particularly when clarifying key boundaries like the red line and bottom line, while using ChatGPT for strategies that are more granular: (P11) \textit{``For example, this work on the red line and bottom line is something that concerns them directly. While for example, which tactics I use for the negotiation is more delegate to me. I would use the normal ChatGPT we saw in these days, more tactically.''} (P11)

\section{Discussion}
\subsection{Negotiators' Hopes and Concerns of Using LLMs in Frontline Negotiation}

\subsubsection{Current and Future Uses of LLMs in Negotiation} Negotiators primarily use ChatGPT for text summarization, quick information retrieval, and reflecting on past negotiations to improve future strategies. They also hope it can anonymize cases to preserve institutional memory and simulate negotiations for training purposes.

\subsubsection{Old Concerns, New Interpretations}
\hfill \break
\textbf{Confidentiality:} The consequences of a confidentiality breach in frontline negotiation are far more serious than in other contexts. Disclosing sensitive details about strategies, stakeholders, or identities could not only undermine the negotiation’s effectiveness but also endanger the safety of the negotiation team. Moreover, negotiators who are not trained in machine learning do not have enough understanding of how LLMs work. They may struggle to assess the risks of using systems like ChatGPT, particularly when dealing with sensitive information. Additionally, this lack of understanding can create a false sense of security, where negotiators might mistakenly believe that the information they input into AI tools is kept confidential. Our findings revealed that as negotiators become more comfortable with ChatGPT over time, they tend to become less cautious about the type of information they share with the tool. This complacency could lead to the unintended disclosure of sensitive information. 

\textbf{Hallucination:} Negotiators believe they have built-in safeguards against the risks of LLM hallucinations. Negotiators rely on several key practices: fact-checking, leveraging  field experience and using tools like LLMs strategically. The more experienced the negotiators are, the less concerned they are about these risks and vice versa. That beng said, because prior studies did indicate that humans often overrely on AI~\cite{bucinca2021trust, inkpen2023complementarity},  future research is needed to determine whether negotiators can truly avoid over-relying on AI.

\textbf{Bias:} In frontline negotiations, the western-centric biases in LLMs raise concerns about their suitability in global contexts. Negotiators worry these biases could limit diversity in perspectives and overlook critical cultural nuances. Attitudes on this issue vary: experienced negotiators are less concerned, seeing bias—whether human or AI-induced—as a manageable aspect of the negotiation process. They feel confident in recognizing and mitigating AI biases, just as they do their own. However, they fear that novice negotiators may struggle to identify and counteract these biases, potentially leading to overreliance on AI outputs. This highlights the need for careful integration of LLMs, with focused training and support for less experienced practitioners.

\subsubsection{New Concerns.}
\hfill \break
\textbf{Mandators' Influence: } 
When integrating AI into their practices, negotiators consider the views and directives of their stakeholders, including the countries or organizations they represent. These stakeholders’ mandates influence whether AI tools can be adopted in negotiation settings. 
Additionally, AI tools for negotiation should help negotiators present and justify their strategies to mandators. These tools should support clear documentation and presentation of processes and outcomes that align with stakeholder expectations. For example, the probe interface generates redlines and bottom lines, which are strategic elements mandators prioritize. 

\textbf{Deploying LLMs Under Resource Limitations: }
Deploying LLMs in resource-limited regions is difficult due to inadequate infrastructure, with limited internet and unstable power disrupting cloud-based and local models. The need for high-performance graphics cards, restricted by export sanctions, further hinders deployment. ~\cite{overclock3d2024, iaps2023}. Unsecured networks in such areas also increase the risk of data breaches, threatening confidentiality~\cite{Journal2022DATAS} .

Therefore, designers should prioritize lightweight, adaptable technology that operates with limited resources, such as decentralized AI models for less powerful hardware. ~\cite{swarm_intelligence_decentralized, distributed_training_llms, gradientcoin_decentralized_llms}.  They can also focus on resource-efficient interventions like negotiation training, which is less time-sensitive and better resourced than active negotiations. LLMs enhance training by creating realistic simulations and teaching specific skills. ~\cite{Shaikh2024Rehearsal, generativeAgents}. Research shows LLMs' potential in simulations, and while they may struggle with complex emotions, they are effective in teaching structured techniques like de-escalation, as outlined in the CCHN handbook(~\cite{CCHN2019}). 

\textbf{LLMs Do Not Raise Questions About Missing Information:} Negotiators noted that LLMs currently cannot identify or inquire about missing information in negotiation files. This limitation is frustrating, as recognizing and addressing gaps is crucial in negotiations. LLMs often respond based on assumptions when data is incomplete, without signaling their lack of full understanding. Unlike LLMs, human negotiators instinctively ask clarifying questions when sensing missing details.

\subsection{Rethinking Interfaces for LLM-Assisted Negotiation}

\subsubsection{Prompt Engineering Challenges.} Existing chat interfaces like ChatGPT are ineffective for tasks requiring specialized skills such as frontline negotiation. Prior research shows that prompt engineering is challenging for most users, ~\cite{Zamfirescu2023Johnny}, and our findings confirm that negotiators are no exception. Many struggled to use the interface effectively, with over half inputting entire case files, expecting comprehensive answers in a single interaction. However, the ChatGPT interface does not naturally support advanced prompt engineering techniques such as Chain of Thought~\cite{wei2023chainofthoughtpromptingelicitsreasoning}, and it inconsistently followed instructions, often neglecting key humanitarian principles.For instance, in a hostage negotiation, it deprioritized rescuing kidnapped workers, and in a student protest case, it failed to consider multiple perspectives. Furthermore, negotiators with less experience and technological fluency faced additional challenges, highlighting the need for AI tools that offer varying levels of guidance and customization to cater to different skill and experience levels. These limitations reveal a mismatch between current LLMs and the demands of negotiation tasks. 

\subsubsection{Bridging the Prompt Skill and Negotiation Experience Gap through Tailored Interfaces.} 

Although the system was not tested in group settings during our study, negotiators offered insights into its potential benefits for team environments. They envisioned the tool aiding in collaborative validation of information, cross-checking data, and identifying agreements and disagreements on negotiation facts and norms. 
Negotiators noted that this tool could enable novice negotiators to create frameworks typically produced only by experienced professionals. One participant explained how the tool could level the playing field, allowing teams to work more efficiently regardless of individual experience levels. 
This suggests that Tailored interfaces grounded in negotiation frameworks could simplify prompt engineering, allowing negotiators to focus on substance rather than technical challenges.

\subsection{Showing the Process rather than Showing the Results}

Current AI systems for negotiation tend to prioritize recommendations and outcomes, which may be effective in transactional settings but fall short in humanitarian negotiations. Negotiators are hesitant to trust AI-generated information without manual verification or confirmation through trusted networks, as their work relies on the accuracy of information and the relationships built through direct communication. This reliance on human verification highlights the unsuitability of goal-centric AI systems for humanitarian negotiation. Tools like Nibble and Practum~\cite{vanhoek2022walmart, nibble2021} , which are designed for transactional negotiation, focus on single recommendations or replacing negotiators, limiting  negotiators' ability to consider alternative options and adapt to changing contexts. Furthermore, these goal-centric systems also push users into backward reasoning, working from a recommended end result, which reduces  cognitive engagement and can lead to overreliance~\cite{zhang2024recommendationsbackwardforwardai}. Moreover, these tools overlook crucial human elements such as empathy and relationship-building, which are vital for navigating complex social environments and fostering cooperation. Negotiators often engage in informal interactions, like multiple tea sessions, to build rapport before tackling core negotiation issues—something AI tools, focused solely on data-driven outcomes, cannot replicate. Therefore, AI's role should be complementary, supporting data processing and scenario analysis while leaving the relational and emotional aspects of negotiation to human expertise.

This distinction between AI's strengths and human expertise reflects the different roles each should play. AI can efficiently process data and simulate scenarios, but the emotional intelligence required for trust-building and empathy remains firmly in the human domain.

\subsection{The Future of LLMs in Frontline Negotiation}

Negotiators who participated in our study believe that AI tools are not, and may never be, essential in humanitarian negotiations, which have been successfully conducted for decades without AI assistance. However, concerns arise if counterparties, such as armed groups, begin using AI, potentially gaining an advantage.  Additionally, given how readily available these tools are\footnote{e.g., OpenAI offers free trial of their currently most powerful model, GPT-4o~\cite{OpenAI2023GPT4}.}, advocacy of not using these tools or even complete ban seems futile. Thus, despite resistance and unresolved concerns, it is crucial to understand the potential impacts of AI on frontline negotiation practices.

AI could assist in tasks such as summarizing positions into readable forms, a task that negotiators frequently perform to clarify and communicate their stance effectively. For example, our probe interface helps distills complex information into concise summaries and support peer validation, and cross-checking of information. Negotiators appreciated these functions as they align with their routine interactions, helping to reinforce existing validation processes and reduce the risk of over-reliance on AI.

\subsection{Towards a Guideline for AI Best Practices for Frontline Negotiators}

Our results showed that currently negotiators and humanitarian organizations are still in early stages of grappling with the effects and impacts of LLMs on negotiation. Organizations like the ICRC have developed an initial set of AI guidelines for negotiation that centers concerns like leaking sensitive information. However, our results indicate that existing guidelines may inadequately address the wide range of potential risks posed by these tools. For example, despite warnings about privacy risks, negotiators may still gradually grow at ease with submitting data to third-party LLM providers. To address these concerns, we recommend that organizations collaborate with technologists to gain a deeper understanding of the technology, establish clear LLM usage guidelines, and provide comprehensive training on the safe use of LLMs in frontline negotiations. There is a need for humanitarian organizations to fully grasp the risks associated with AI before deploying these tools in the field. Because of the complexity of these tools, technologists should actively assist organizations in understanding AI's capabilities and limitations. Conversely, technologists need to learn from frontline negotiators and use these learnings in designing future technologies. Finally, to bridge the knowledge gap, proper training plays an essential role. Frontline Associates already actively organize workshops and communities of practice to foster a deeper understanding of these tools among negotiators, a practice that should be continued and expanded.

\begin{acks}

\end{acks}

\bibliographystyle{ACM-Reference-Format}
\bibliography{sample-base, zilinma, new_bibs}

\appendix
\section{Negotiation Tools}
\label{apx:negotiation_tools}
\subsection{Island of Agreement (IoA)}
\begin{description}
\item[Islands of Agreement] This concept refers to the areas of common ground or shared understanding between parties in a negotiation, despite their overall disagreements. It is based on the paradox that for any disagreement to exist, there must also be some level of agreement or shared perspective. The Islands of Agreement serve as a starting point for dialogue and building trust.
\end{description}
\begin{example}[Access to IDP Camps]
In a negotiation between Food Without Borders (FWB) and the Governor of a district regarding access to IDP camps:
\begin{itemize}
\item \textbf{Agreed Facts:}
\begin{itemize}
\item There are displaced persons from Country A in the no man's land.
\item People are blocked in the no man's land, in a difficult situation in terms of shelter and nutrition.
\item There is little prospect of improvement without immediate access to the displaced.
\end{itemize}
\item \textbf{Convergent Norms:}
\begin{itemize}
\item There is a legitimate border between Country A and Country B. B has the right to defend the integrity of its territory and prevent illegal entry.
\item We should not allow people to die from starvation.
\item People have a right to flee armed violence.
\end{itemize}
\item \textbf{Contested Facts:}
\begin{itemize}
\item The exact number of displaced persons in the area.
\item The severity of the situation and who is in most urgent need.
\item The presence of armed elements among the civilians.
\end{itemize}
\item \textbf{Divergent Norms:}
\begin{itemize}
\item Whether humanitarian organizations have a right of access to people in need under international law.
\item Whether people have a right to enter Country B simply because they flee armed violence.
\item The priority of government security concerns versus humanitarian needs.
\end{itemize}
\end{itemize}
\end{example}

\subsection{Iceberg and Common Shared Space (CSS)}
\begin{figure}[H]
  \centering
  \includegraphics[width=0.6\linewidth]{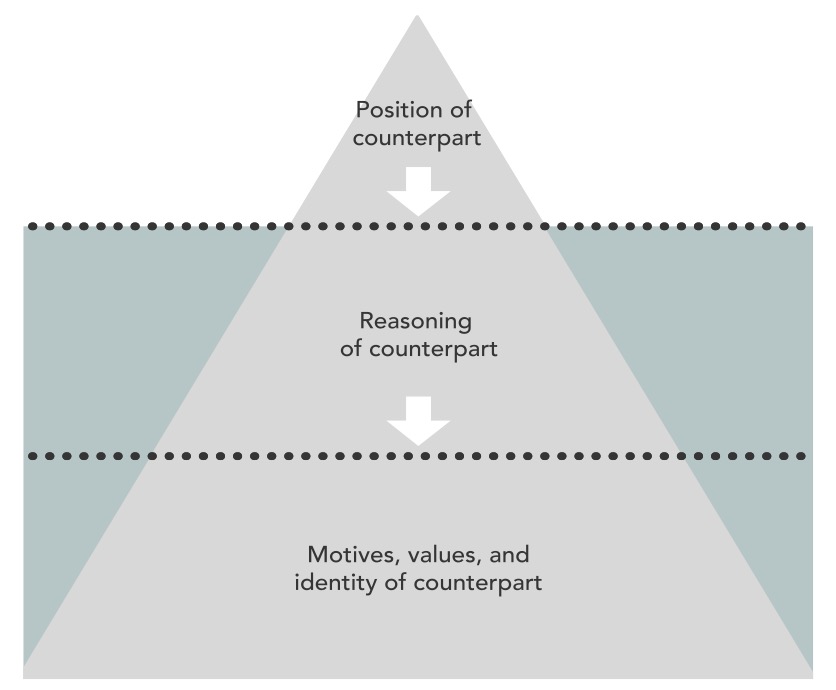}
  \caption{Graphic Example of Iceberg and CSS Framework~\cite{CCHN2019}}
  \label{fig:iceberg}
\end{figure}

\begin{description}
\item[Iceberg Model] This model is used to analyze the position of both the humanitarian organization and the counterpart in a negotiation. It consists of three levels (Also see figure~\ref{fig:iceberg}):
\begin{enumerate}
\item \textbf{Position (WHAT):} The visible stance or demand at the top of the iceberg.
\item \textbf{Reasoning (HOW):} The logic and interests supporting the position.
\item \textbf{Values and Motives (WHY):} The underlying principles, needs, and drivers at the base of the iceberg.
\end{enumerate}
\item[Common Shared Space (CSS)] This refers to the area of potential agreement between parties in a negotiation, where their interests, reasoning, and values overlap or can be reconciled.
\end{description}

\begin{example}[Health for All (HfA) Negotiation]
HfA is negotiating with tribal leaders over a hospital closure and staff detention.
\textbf{HfA's Iceberg:}
\begin{itemize}
\item \textbf{Position (WHAT):} Immediate release of staff and evacuation from District A.
\item \textbf{Reasoning (HOW):} Ensure staff safety, scale down surgical activities, hand over hospital to third party.
\item \textbf{Values (WHY):} Humanitarian principles, duty of care, professional health standards.
\end{itemize}
\textbf{Tribal Leaders' Iceberg:}
\begin{itemize}
\item \textbf{Position (WHAT):} Keep hospital fully operational under HfA or equivalent.
\item \textbf{Reasoning (HOW):} Maintain employment for guards, compensate families of injured/deceased guards.
\item \textbf{Values (WHY):} Community welfare, tribal loyalty, economic stability.
\end{itemize}
\textbf{Common Shared Space (CSS):}
\begin{itemize}
\item Shared concern for community health.
\item Recognition of guards' service and sacrifices.
\item Desire to maintain reputation and relationships.
\item Need for evidence-based decision-making on health needs.
\item Importance of addressing emergency medical needs.
\item Necessity of community consultation in healthcare planning.
\end{itemize}

\end{example}

\subsection{Redline and Bottomline}

\begin{description}
\item[Red lines] Red lines are defined as the outer limits of the possible areas of agreement. They set the parameters within which parties to the negotiation must remain while attempting to maximize their shared benefit. Red lines are generally specified in the mandate given to the negotiator and informed by applicable laws and institutional policies. Red lines cannot be crossed, as doing so would have significant consequences regarding the validity and legality of the agreement and may impact the legitimacy of the negotiator and organization. The negotiator is not allowed to set or revise red lines.
\item[Bottom lines] Bottom lines are a tactical tool used by the negotiator to set limits to the conversation when options under consideration show rising risks and diminishing benefits. Bottom lines are under the control of the negotiator as a means to suspend or postpone consideration of additional options below a certain threshold. Before considering options beyond the bottom line, the negotiator may consult again with their hierarchy. The results of this consultation may impact the location of the bottom line.
\end{description}

\begin{example}[Food Without Borders Negotiating Access to IDP Camp]
Food Without Borders (FWB) is negotiating access to an IDP camp to distribute food rations. The negotiation involves several key points:
\begin{description}
\item[Ideal outcome (Point A):] All food rations are distributed only to the affected IDP population based on their nutritional needs. FWB can hire and pay in cash the day laborers of their choice to assist in its work in the IDP camp.
\item[Bottom line (Point B):] Food rations should be limited to IDPs but are not necessarily dependent on their individual nutritional needs. FWB could consider including family members of local guards in need as part of the food distribution process, even though they are not recognized as formal IDPs. Direct distribution to the local guards, however, is not permitted.
\item[Red line (Point C):] FWB can only distribute food rations to the IDP population and other people in need. It cannot use the food rations as a means of payment for laborers. It further cannot provide any direct assistance to armed personnel.
\end{description}
In this example:
\begin{itemize}
\item The bottom line (B) represents the point where FWB negotiators feel the compromises are reaching their limit, but they can still make decisions within their mandate.
\item The red line (C) is set by FWB's institutional policy prohibiting the use of food rations as currency or compensation for labor, which the negotiator cannot cross without referring back to higher authorities.
\end{itemize}
\end{example}

\section{Survey questions}
\label{apx:survey}
\begin{enumerate}
    \item \textbf{How many years of negotiation experience do you have?} 
    \item \textbf{How often do you use a computer for work?}
    \begin{itemize}
        \item Once a week or less
        \item A few times a week
        \item A couple of hours most days
        \item Many hours on most days
    \end{itemize}
    \item \textbf{What is the highest level of education you have received or are pursuing?}
    \begin{itemize}
        \item Pre-high school
        \item High school
        \item College
        \item Masters or professional degree
        \item PhD
    \end{itemize}
    \item \textbf{How often do you use AI?}
    \begin{itemize}
        \item Once a week or less
        \item A few times a week
        \item A couple of hours most days
        \item Many hours on most days
    \end{itemize}
    \item \textbf{If you said yes to the question before, how often do you use AI for work relating to frontline negotiation?}
    \begin{itemize}
        \item Once a week or less
        \item A few times a week
        \item A couple of hours most days
        \item Many hours on most days
    \end{itemize}
\end{enumerate}

If you are interested in a 1-hour interview to help us learn more about AI in negotiations, please share your email and schedule an interview here: 

\section{Interview Questions: Formative Study}
\label{apx:formative}
\subsection{Part 1: Semi-Structured Interviews with Negotiators}

The following questions focus on the negotiation process:
\begin{itemize}
    \item Tell me about how you prepare a case.
    \item What was the most difficult part of such a process?
    \item How do you collaborate with colleagues?
\end{itemize}

\subsection{Part 2: Demo of ChatGPT Generating Iceberg CSS}

Demo of an automated system that creates negotiation visualizations. Questions asked after the demo:
\begin{itemize}
    \item What did you like about the tool?
    \item Do you think you can use this tool in your workflow?
    \item Do you have any concerns about using this tool?
    \item Do you have anything else that you want to mention?
\end{itemize}

\section{Interview Questions: Probe Study}
\label{apx:probe}
Each question is asked for both systems (ChatGPT and probe interface) when time permits.

\subsection{1. Impressions of the System}
\begin{itemize}
    \item What was your impression of this system?
\end{itemize}

\subsection{2. Added Value of the System}
\begin{itemize}
    \item What was the greatest added value of this system?
\end{itemize}

\subsection{3. Problems and Improvements}
\begin{itemize}
    \item What did you find problematic or in need of improvement?
\end{itemize}

\subsection{4. Desired AI Features}
\begin{itemize}
    \item What other AI features would you like to see in these systems?
\end{itemize}

\subsection{5. Early-Stage Technology Feedback}
\begin{itemize}
    \item Where else do you think AI could be added to your process? 
    \item Do you see the value of this technology in your work?
\end{itemize}

\subsection{6. Future Development Expectations}
\begin{itemize}
    \item How disappointed would you be if we never developed anything AI-related with you through this collaboration?
    \item How disappointed would you be if no one ever made technology like this real? Why?
\end{itemize}
\end{document}